\pdfoutput=1

\documentclass[11pt]{article}

\usepackage[]{acl}

\usepackage{times}
\usepackage{latexsym}
\usepackage{amssymb}
\usepackage[T1]{fontenc}
\usepackage[utf8]{inputenc}

\usepackage{microtype}

\usepackage{graphicx}
\usepackage{multirow}
\usepackage{adjustbox}
\usepackage{booktabs}
\usepackage{array}
\usepackage{caption}
\usepackage{subcaption}
\usepackage{svg}
\usepackage{forest}

\usepackage{xcolor,colortbl}

\usepackage{soul}
\definecolor{hlgreen}{HTML}{B2D5CB}
\definecolor{hlblue}{HTML}{ADD8E6}
\definecolor{hlyellow}{HTML}{EADDCA}

\usepackage{xspace}

\DeclareSymbolFont{extraup}{U}{zavm}{m}{n}
\DeclareMathSymbol{\varheart}{\mathalpha}{extraup}{86}

\title{AudioBench: A Universal Benchmark for Audio Large Language Models}

\author{
\textbf{
Bin Wang\textsuperscript{$\diamondsuit$} 
\ \quad 
Xunlong Zou\textsuperscript{$\diamondsuit$} 
\ \quad 
Geyu Lin\textsuperscript{$\diamondsuit$} 
\ \quad 
Shuo Sun\textsuperscript{$\diamondsuit$} 
\ \quad 
Zhuohan Liu\textsuperscript{$\diamondsuit$}} \\
\textbf{
Wenyu Zhang\textsuperscript{$\diamondsuit$} 
\ \quad 
Zhengyuan Liu\textsuperscript{$\diamondsuit$} 
\ \quad 
AiTi Aw\textsuperscript{$\diamondsuit$} 
\ \quad 
Nancy F. Chen\textsuperscript{$\diamondsuit,\dag$}} \\
\textsuperscript{$\diamondsuit$}Institute for Infocomm Research (I$^2$R), A*STAR, Singapore\\
\textsuperscript{$\dag$}Centre for Frontier AI Research (CFAR), A*STAR\\
\texttt{wang\_bin@i2r.a-star.edu.sg} \\
}
  
\begin{document}
\maketitle

\begin{abstract}

    We introduce AudioBench, a universal benchmark designed to evaluate Audio Large Language Models (AudioLLMs). It encompasses 8 distinct tasks and 26 datasets, among which, 7 are newly proposed datasets. The evaluation targets three main aspects: speech understanding, audio scene understanding, and voice understanding (paralinguistic). Despite recent advancements, there lacks a comprehensive benchmark for AudioLLMs on instruction following capabilities conditioned on audio signals. AudioBench addresses this gap by setting up datasets as well as desired evaluation metrics. Besides, we also evaluated the capabilities of five popular models and found that no single model excels consistently across all tasks. We outline the research outlook for AudioLLMs and anticipate that our open-sourced evaluation toolkit, data, and leaderboard will offer a robust testbed for future model developments.\footnote{\url{https://github.com/AudioLLMs/AudioBench}}


    
\end{abstract}

\section{Introduction}

    Foundation models, built upon large language models (LLMs), have demonstrated strong capability in handling diverse tasks and diverse modalities~\cite{bommasani2021opportunities,achiam2023gpt,team2023gemini}. There have been a series of benchmarks proposed recently to provide a holistic evaluation of LLMs~\cite{hendryckstest2021,cobbe2021gsm8k,wang2024seaeval,rein2023gpqa}, image-enhanced multimodal LLMs~\cite{marino2019ok,yue2023mmmu,padlewski2024vibe}, and even video-enhanced multimodal LLMs~\cite{xiao2021next,li2023mvbench}. However, the audio large language models (AudioLLMs) lag, and the performance of their audio interpretation capability is often intractable and not systematically compared in different tasks. The existing evaluation regime does not cover the breadth of their possible use cases.

    Although instruction-following audio-language models have attracted significant interest from both industry and research communities, their reported evaluation datasets are notably different. For instance, the Qwen-Audio-Chat~\cite{chu2023qwen} model has tested on 12 datasets, whereas the SALMONN~\cite{tang2023salmonn} model used 15 datasets, with only two datasets in common. Although WavLLM~\cite{hu2024wavllm} has been compared with some earlier models, the scope of evaluation tasks remains limited, which makes the comparison less justified. Another limitation arises from the evaluation set; similar to text-based LLMs~\cite{chiang2024chatbot}, evaluations predominantly rely on previous datasets and metrics~\cite{hendryckstest2021}. Given that these models are expected to respond flexibly following the instructions, conventional evaluation metrics are not well prepared for this purpose. Consequently, it is crucial to develop innovative benchmarks for AudioLLMs. Since their ultimate goal is to handle diverse audio inputs and respond accurately to user queries, the evaluation set should include not only traditional speech tasks but also new datasets and metrics that better reflect their particular nature and assess the instruction-following capability of models, such as zero-shot speech question answering, etc.

    \begin{figure*}[t]
         \centering
             \includegraphics[width=1.00\textwidth]{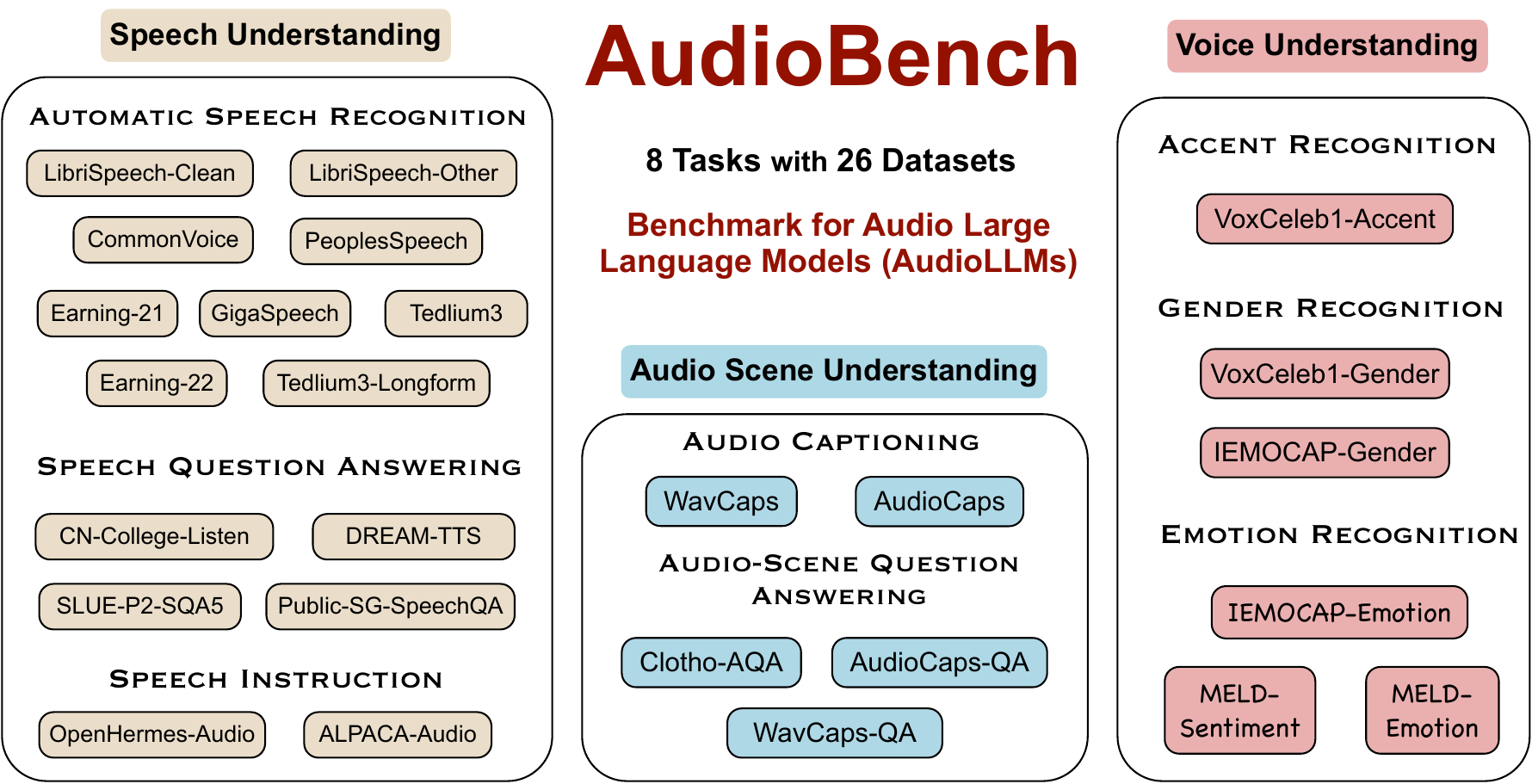}
            \caption{Overview of \textbf{AudioBench} datasets.}
            \label{fig:overview}
    \end{figure*}
    
    In this study, we introduce AudioBench as a comprehensive evaluation benchmark specifically designed for general instruction-following Audio Large Language Models, which we refer to as AudioLLMs. These models are designed to comprehend general audio inputs and user queries. The tasks should include but are not limited to, understanding speech semantics, interpreting environmental sounds, and recognizing paralinguistic features, as well as combinations of these elements. Therefore, as shown in Figure~\ref{fig:overview}, we incorporate 8 tasks and 26 datasets covering these aspects, with 7 of the datasets being newly adapted or collected to address gaps in the existing datasets.

    For a more comprehensive assessment of the models, our benchmark incorporates a variety of prompt templates to evaluate model compatibility with different instructions. This is because we found that some models are more robust to seen instruction while do not generalize to unseen ones which could lead to robustness issues for deployment. Additionally, we vary the input lengths from seconds to minutes to better assess model performance on longer audio sequences. Our test suite includes 400+ hours of audio, distributed across 100k+ samples. Our \textbf{AudioBench} toolkit is available, enabling the evaluation of future AudioLLMs by simply implementing a customized model inference program.

    We also explored and decided what evaluation metrics to leverage, particularly for open-ended generation of AudioLLMs. Inspired by text-based and vision-based large language models (LLMs), which can be accessed with multiple-choice and open-ended questions, AudioLLMs are designed to manage diverse instruction and need specific metrics to cover the unconstrained generation space. This introduces difficulties in quantitatively assessing their outputs. Most practices go to model-as-judge methods~\cite{padlewski2024vibe,yang-etal-2024-air}. However, cloud-sourced models, e.g. GPT-4, can be costly and subject to instability introduced by commercial version updates. Therefore, we investigate open-source model-as-judge methods and validate the best practices by correlation studies. Our findings indicate that LLaMA-3-70B-Instruct model demonstrates a higher correlation compared to judgment models like Prometheus-2~\cite{kim2024prometheus} which is specifically fine-tuned for evaluation purposes.
    
    Finally, we evaluated 5 models to offer a comprehensive review of various solutions, including both end-to-end and cascaded models. Our benchmark results indicate that none of the models performed exceptionally well across all criteria, highlighting significant opportunities for future advancements in model capabilities. To guide these developments, we also include an in-depth discussion on future challenges and directions.

\section{Related Work}

    \textbf{Multimodal LLM Benchmark.} Recent progress in both text-based and multimodal LLMs has spurred the creation of a series of benchmarks designed to monitor and direct their development efficiently. Our research is consistent with earlier benchmarks on multimodal LLMs, primarily concentrating on visual components. There is a lack of comprehensive benchmarks for multimodal LLMs that provide a holistic understanding of audio inputs including speech, ambient sounds, paralinguistic features, etc. When classifying the models targeted for evaluation, we can categorize them into several distinct types: Text-LLMs, Vision-LLMs, and Audio-LLMs.
        
    Text-based LLMs present the most fundamental advancements in the past years and serve as the core for developing multimodal systems. Their evaluation is also multifaceted covering aspects such as reasoning~\cite{zellers2019hellaswag,cobbe2021training,shi2022language}, subject knowledge~\cite{clark2018think,hendryckstest2021,huang2024c}, safety~\cite{zhang2023safetybench,li2024salad}, multilingual capability~\cite{wang2024seaeval,ahuja2023mega}, and more~\cite{chiang2024chatbot,zheng2024judging,dubois2024length}. Text-based LLMs have driven fundamental advancements in recent years~\cite{jiang2023mistral,achiam2023gpt,gemma} and have significantly contributed to the evolution of vision LLMs that integrate visual understanding with text queries~\cite{team2023gemini}. Additionally, a series of multimodal benchmarks specifically tailored for vision LLMs have also been established.

    The evaluation of Vision-LLMs is multifaceted, encompassing aspects such as multidisciplinary understanding~\cite{yue2023mmmu,zhang2024m3exam,hu2024omnimedvqa}, perception tests~\cite{yu2023mm,liu2023mmbench}, mathematical reasoning~\cite{li2024multimodal,zhang2024mathverse}, and video understanding~\cite{li2023mvbench,ning2023video,liu2024tempcompass,fu2024video}. Notably, video understanding benchmarks assess modalities including both visual and audio elements. However, these benchmarks cannot be directly applied to audio understanding, as they predominantly rely on visual inputs with audio serving merely as a supplementary feature. Consequently, there is an urgent need to standardize benchmarks for AudioLLMs that specifically measure model performance with audio context. Such benchmarks are essential for advancing research and improving the capabilities of models that focus on audio processing and understanding.

    \textbf{Advancements in AudioLLMs}. Although no comprehensive evaluation benchmarks currently exist, we observe various multimodal foundation models that enhance speech and audio understanding, utilizing either cascaded methods or integrated multitask optimization. Some models specialize in Automatic Speech Recognition (ASR)~\cite{shu2023llasm,rao2024emergent} or other non-speech audio tasks~\cite{Kong2024AudioFlamingo}. However, ideal speech-text foundation models often employ multitasking approaches. Notable examples include AudioGPT~\cite{huang2023audiogpt}, SpeechGPT~\cite{zhang2023speechgpt}, SALMONN~\cite{tang2023salmonn}, Qwen-Audio~\cite{chu2023qwen}, AudioPALM~\cite{rubenstein2023audiopalm}, LTU-AS~\cite{gong2023ltuas}, VioLA~\cite{wang2023viola}, LLaSM~\cite{shu2023llasm}, Pengi~\cite{deshmukh2023pengi}, WavLLM~\cite{hu2024wavllm}, UniverSLU~\cite{arora2024universlu}, SpeechVerse~\cite{Das2024SpeechVerse}, and Qwen2-Audio~\cite{chu2024qwen2} each focusing on a subset of tasks tailored to their specific use cases. In this study, we focus on four models—WavLLM, Qwen-Audio-Chat, SALMONN, and Qwen2-Audio-Instruct—that represent a broader range of task-handling capabilities.

\section{AudioLLM Benchmark (AudioBench)}

    \subsection{Core Concept}

        The design of AudioLLMs is to interpret audio content and respond flexibly to user queries, handling tasks from basic speech recognition to complex reasoning. To evaluate these models effectively, we should assess their ability to understand not only speech but also environmental sounds and paralinguistic features such as emotion, gender, and accent. Therefore, our evaluation involves designing benchmarks that cover three key aspects: 1) speech understanding, focusing on the interpretation of the semantic content within the audio; 2) audio scene understanding, focusing on non-human audio sounds; and 3) voice understanding, focusing on non-speech but human-related information and including paralinguistic tasks such as emotion, accent, and gender.         For each task, we compile existing datasets and develop new ones tailored to the requirements for comprehensive understanding and reasoning. Figure~\ref{fig:overview} is an illustration of the tasks and datasets used in our setup.

        In AudioBench, the focus is on audio understanding of English semantics, where most models have not yet demonstrated satisfactory results. Meanwhile, multilingual understanding, which includes speech translation and code-switching, represents a useful application that could be integrated with future work.

    \begin{table*}[htb]
        \centering
        \begin{adjustbox}{width=0.95\textwidth,center}
        \begin{tabular}{ | c | l | c | c | c | c |}
        \toprule
         \textbf{Category} & \textbf{Dataset Name} & \textbf{\#Samples} & \textbf{Hours} & \textbf{Avg.L/Min.L/Max.L(s)} & \textbf{Metrics} \\  \hline 

        \multicolumn{6}{|c|}{\sethlcolor{hlyellow}\hl{\textbf{Speech Understanding}}}
            \\ \hline
         
         \multirow{9}{*}{ASR} & LibriSpeech-Clean   & 2.6k   & 5.40 & 7.43 / 1.28 / 34.96 & \multirow{9}{*}{WER\textsubscript{$(\downarrow)$}} \\
         & LibriSpeech-Other   & 2.9k   & 5.34 & 6.55 / 1.47 / 34.51 & \\
         & CommonVoice-15      & 16k  & 26.95 & 5.93 / 1.34 / 105.67 & \\
         & PeoplesSpeech       & 32k  & 59.20 & 6.54 / 1.00 / 99.91 & \\
         & GigaSpeech          & 18k  & 35.09 & 6.77 / 1.00 / 22.03 & \\
         & Tedlium3            & 1.1k   & 2.61 & 8.24 / 1.07 / 32.55 & \\
         & Tedlium3-Longform   & 11      & 2.61 & 0.8k / 0.3k / 1.6k & \\ 
         & Earning-21          & 44      & 39.26 & 3.2k / 1k / 5.7k & \\
         & Earning-22          & 125     & 119.88 & 3.4k / 0.87k / 7.4k & \\ \hline
         
         \multirow{4}{*}{SQA} & \cellcolor{yellow!15}CN-College-Listen   & 2.2k & 13.3 & 21.09 / 5.76 / 137.82 & \multirow{4}{*}{M.J.\textsubscript{$(\uparrow)$}} \\     
         & SLUE-P2-SQA5 & 408 & 4.5 & 39.85 / 13.00 / 40.0 & \\         
         & \cellcolor{yellow!15}DREAM-TTS  & 1.9k & 18.1 & 34.14 / 3.14 / 261.85 & \\
         & \cellcolor{yellow!15}Public-SG-SpeechQA  & 688 & 7.6 & 39.86 / 15.78 / 95.71 & \\ \hline

         \multirow{2}{*}{SI} & \cellcolor{yellow!15}OpenHermes-Audio    & 100 & 0.16 & 5.95 / 2.04 / 15.77 & \multirow{2}{*}{M.J.\textsubscript{$(\uparrow)$}} \\         
         & \cellcolor{yellow!15}ALPACA-Audio & 100 & 0.12 & 4.32 / 1.80 / 8.85 & \\  \hline

        \multicolumn{6}{|c|}{\sethlcolor{hlblue}\hl{\textbf{Audio Scene Understanding}}}
            \\ \hline

         \multirow{3}{*}{AQA} & Clotho-AQA  & 2.2k & 14.1 & 22.59 / 15.03 / 29.97 & \multirow{3}{*}{M.J.\textsubscript{$(\uparrow)$}} \\         
         & \cellcolor{yellow!15}WavCaps-QA  & 304 & 0.87 & 10.28 / 1.0 / 30.62 & \\         
         &  \cellcolor{yellow!15}AudioCaps-QA & 313 & 0.86 & 9.86 / 3.27 / 10.00 & \\  \hline

         \multirow{2}{*}{AC} & WavCaps     & 1.7k & 4.9 & 10.22 / 1.00 / 30.97 & \multirow{1}{*}{M.J.\textsubscript{$(\uparrow)$} \&} \\         
         & AudioCaps   & 4.4k & 12.1 & 9.86 / 1.74 / 10.0 & \multirow{1}{*}{METEOR\textsubscript{$(\uparrow)$}} \\  \hline

        \multicolumn{6}{|c|}{\sethlcolor{pink}\hl{\textbf{Voice Understanding}}}
            \\ \hline

         \multirow{3}{*}{ER} & IEMOCAP-Emotion & 1k & 1.3 & 4.51 / 0.75 / 24.12 & \multirow{3}{*}{M.J.\textsubscript{$(\uparrow)$}} \\         
         & MELD-Sentiment & 2.6k & 2.4 & 3.35 / 0.13 / 304.9 & \\         
         & MELD-Emotion  & 2.6k & 2.4 & 3.35 / 0.13 / 304.9 & \\  \hline

         AR & VoxCeleb1-Accent & 4.8k & 11.2 & 8.27 / 3.96 / 69.04 & \multirow{1}{*}{M.J.\textsubscript{$(\uparrow)$}} \\ \hline

         \multirow{2}{*}{GR} & VoxCeleb1-Gender & 4.8k & 11.2 & 8.27 / 3.96 / 69.04 & \multirow{2}{*}{M.J.\textsubscript{$(\uparrow)$}} \\         
         & IEMOCAP-Gender & 1k & 1.26 & 4.55 / 0.69 / 26.77 &  \\ 
    
        \bottomrule
        \end{tabular}
        \end{adjustbox}
        \caption{
        Statistics of AudioBench Dataset Statistics. \colorbox{yellow!15}{Yellow} refers to our newly expanded and collected datasets to accommodate the missing of suitable datasets. WER refers to Word-Error-Rate. METEOR is a common metric for audio captioning. M.J. refers to model-as-judge, where we deploy Llama-3-70B-Instruct in the current setup. 
        } 
        \label{tab:statistics}
    \end{table*}

    \subsection{Evaluation Setup}

        From a use case perspective, we expect LLMs to respond to user queries and generate natural responses. Therefore, we expect the model's response to adhere to user queries and the generation style can adapt accordingly. Most conventional speech models and tasks have a constrained output space. Given classification tasks as an example, output is typically restricted to a finite set of categories or classes. However, the generation space of AudioLLMs is much broader and more complex, which could lead to difficulties in evaluation.
        
        Therefore, we employed the Model-as-Judge (M.J.) approach for most tasks except for automatic speech recognition, where word error rate (WER) is used as the sole metric, and audio captioning, where the METEOR score is utilized as an additional measure. In subsequent experiments, we thoroughly validated the effectiveness of our proposed evaluation approach. We rescaled the M.J. scores to a 100-point scale for easier comparison.
        

        Besides, we found that models can be less robust to diverse instruction, which can negatively impact their applicability and overall user experience. This issue is also present in pioneering text-based LLMs~\cite{zhu2023promptbench,wang2024seaeval} but it is even more severe in AudioLLMs due to the complexity of processing multiple modalities during the fusion process. Therefore, we propose to leverage multiple instructions to evaluate a dataset, especially when the initial query lacks diversity. This approach is particularly relevant for tasks such as ASR, ER, AR, GR, and AC. The detailed analysis is presented in Section~\ref{sec:results_analysis}.

    \subsection{Tasks}

        In this section, we discuss the functionality and applicability of each task within this benchmark. For newly proposed datasets, we provide a detailed description of their development process. The detailed statistics are provided in Table~\ref{tab:statistics}.

        \subsubsection{Speech Understanding}

        \textbf{Automatic Speech Recognition (ASR)}.
        As a fundamental task in speech processing, ASR aims to convert spoken content into text format. It measures the accuracy of speech-to-text conversion and requires robust algorithms capable of understanding and processing a wide range of linguistic nuances and dialects. ASR systems must be equipped to handle diverse environments, speech rates, accents, and background noises, ensuring that they deliver high-quality transcription under various conditions. To test these capabilities, we included 9 datasets, 3 of which contain audio in a long-form format. Current AudioLLMs struggle with long audio files, which may exceed 10 minutes in duration, presenting a potential area for enhancement in future models. For our current evaluations, we segment the long audio into smaller chunks and then reassemble them for assessment if the model naively does not support the original length. Examples can be found at Table~\ref{example:asr}.

    \begin{table*}[htb]
        \centering
        \begin{adjustbox}{width=1.00\textwidth,center}
        \begin{tabular}{ l | c | c | c | c | c }
        \toprule
         \multirow{2}{*}{\textbf{Dataset Name}} & \multicolumn{4}{c|}{\textbf{AudioLLMs}} & \multirow{2}{*}{\textbf{Whisper+Llama3}} \\ \cline{2-5}
         & \textbf{SALMONN} & \textbf{Qwen-Audio-Chat} & \textbf{WavLLM} & \textbf{Qwen2-Audio-Instruct} \\\hline 

        \multicolumn{5}{l}{\sethlcolor{hlyellow}\hl{\textbf{Speech Understanding}}}
        \\ \hline

         \emph{LibriSpeech-Clean}\textsubscript{$(\downarrow)$}  & 55.58 & 2.25   & \textbf{2.10}  & 3.20 & 1.83 \\
         \emph{LibriSpeech-Other}\textsubscript{$(\downarrow)$}  & 41.80 & \textbf{4.16} & 4.80  & 6.07 & 3.71 \\
         \emph{CommonVoice-15}\textsubscript{$(\downarrow)$}     & 33.75 & 11.65  & 14.53 & \textbf{11.44} & 9.89 \\
         \emph{PeoplesSpeech}\textsubscript{$(\downarrow)$}      & 34.33 & 30.72  & 37.92 & \textbf{22.32} & 14.54 \\
         \emph{GigaSpeech}\textsubscript{$(\downarrow)$}         & 14.22 & 13.32  & 15.49 & \textbf{11.89} & 9.51 \\
         \emph{Tedlium3}\textsubscript{$(\downarrow)$}           & 8.56  & \textbf{4.00}   & 6.62  & 6.39 & 3.81 \\
         \emph{Tedlium3-Longform}\textsubscript{$(\downarrow)$}  & \textbf{18.39} & 45.29  & 45.37 & 95.35 & 4.75 \\ 
         \emph{Earning-21}\textsubscript{$(\downarrow)$}         & \textbf{26.87} & 38.46  & 64.47 & 98.65 & 11.77 \\
         \emph{Earning-22}\textsubscript{$(\downarrow)$}         & \textbf{36.38} & 51.18  & 66.72 & 98.84 & 15.61 \\
         \hline
         
         \emph{CN-College-Listen}  & 50.51 & 60.85 & 65.43 & \textbf{74.50} & 85.25 \\     
         \emph{SLUE-P2-SQA5}       & 78.24 & 76.12 & \textbf{83.92} & 80.05 & 82.99 \\         
         \emph{DREAM-TTS}          & 55.93 & 57.76 & 64.56 & \textbf{66.70} & 86.09 \\         
         \emph{Public-SG-SpeechQA} & 56.77 & 57.47 & \textbf{58.55} & 58.31 & 64.94 \\ 

         \emph{OpenHermes-Audio}   & 19.20 & 11.00 & 22.40 & \textbf{44.80} & 63.0 \\         
         \emph{ALPACA-Audio}       & 12.40 & 9.60 & 21.60 & \textbf{52.60} & 70.8 \\  \hline

        \multicolumn{3}{l}{\sethlcolor{hlblue}\hl{\textbf{Audio Scene Understanding}}}
            \\ \hline

         \emph{Clotho-AQA}     & 51.18 & \textbf{58.20} & 43.01 & 50.92 & 29.47 \\         
         \emph{WavCaps-QA}     & \textbf{46.25} & 38.68 & 26.25 & 44.47 & 17.38 \\         
         \emph{AudioCaps-QA}   & 47.03 & \textbf{47.99}  & 29.84 & 45.75 & 16.71 \\ 

         \emph{WavCaps}\textsubscript{(M.J.)} & 21.16 & 29.25 & 6.40 & \textbf{33.78} & 3.45 \\           
         \emph{AudioCaps}\textsubscript{(M.J.)} & 34.37 & \textbf{47.99} & 4.17 & 40.78 & 2.47 \\  

         \emph{WavCaps}\textsubscript{(METEOR)} & 17.72 & \textbf{24.02} & 9.78 & 21.34 & 13.89 \\        
         \emph{AudioCaps}\textsubscript{(METEOR)} & 21.20 & \textbf{27.70} & 6.70 & 19.89 & 7.95 \\  

         \hline

        \multicolumn{3}{l}{\sethlcolor{pink}\hl{\textbf{Voice Understanding}}}
            \\ \hline

         \emph{IEMOCAP-Emotion}  & 21.56 & 27.34 & 45.91 & \textbf{49.30} & 34.43 \\         
         \emph{MELD-Emotion}     & 33.06 & \textbf{50.57} & 41.07 & 40.54 & 33.36 \\  
         \emph{MELD-Sentiment}   & 41.87 & 43.87 & 50.08 & \textbf{53.49} & 43.87 \\

         \emph{VoxCeleb1-Accent} & 28.06 & \textbf{45.70} & 37.65 & 29.19 & 39.33 \\ 

         \emph{VoxCeleb1-Gender} & 88.90 & 70.56 & 70.51 & \textbf{99.12} & 53.41 \\         
         \emph{IEMOCAP-Gender}   & \textbf{51.60} & 51.13 & 45.29 & 49.30 & 51.50 \\ 
    
        \bottomrule
        \end{tabular}
        \end{adjustbox}
        \caption{Main results of four AudioLLMs and one cascade model. The word-error-rate (WER) for ASR tasks is the lower the better\textsubscript{$(\downarrow)$}.}
        \label{tab:results}
    \end{table*}
    
        \noindent\textbf{Speech Question Answering (SQA)}. The task involves responding to questions based on speech-related audio content. However, there is a lack of suitable datasets. In this part, we involve both monologue understanding and dialogue understanding tasks. Three datasets are newly curated. Examples can be found in Table~\ref{example:sqa}.

        First, we collect questions from the English listening comprehension section of China's national college entrance examinations, aimed at assessing students' ability for listening comprehension in both academic and everyday situations. We manually compiled 271 questions and integrated them with 2000 questions sourced from \citet{hu2024wavllm}, which includes questions from approximately 130 different exam sets. In line with the free-text QA format, we presented the questions without multiple-choice options. The dataset is named as \emph{CN-College-Listen}.

        The second dataset is DREAM-TTS, which is built upon text-based dialogue comprehension datasets DREAM~\cite{sun2019dream}. We leverage one SOTA TTS engine~\cite{casanova2024xtts} to convert the text input into the spoken format. We leverage 60 speakers in the generation and the genders are kept consistent with the dialogue content.

        For Public-SG-SpeechQA dataset, we chose four public speaking videos from Singaporean politicians, accompanied by clean transcriptions. These transcriptions were manually segmented, and five questions per segment were generated using LLMs. Each question and its corresponding reference answer were manually reviewed to ensure their validity, where around 30\% of the samples were discarded. Ultimately, this process resulted in the collection of 688 speech question-answer pairs for evaluation.

        \noindent\textbf{Speech Instruction (SI)}. This task evaluates whether the model can directly follow instructions provided through speech input, mirroring a natural human-computer interaction. Specifically, the question is delivered via audio, and the model is expected to understand and generate responses in the appropriate text format. This approach offers unique benefits. For instance, by incorporating paralinguistic information such as emotions, the model can adapt its responses based on the user's emotional state. However, finding suitable testing examples has been a challenge. In this study, we synthesized audio instruction~\cite{casanova2024xtts} from existing instruction-following datasets: ALPACA~\cite{alpaca} and OpenHermes~\cite{OpenHermes}. We then involve humans in selecting instances to ensure that 1) the speech is accurately synthesized and 2) the content is suitable as spoken instructions. As a result, around 10.5\% samples are kept and the examples can be found in Table~\ref{example:si}.
        
    \begin{figure*}[htb]
     \centering
         \includegraphics[width=1.00\textwidth]{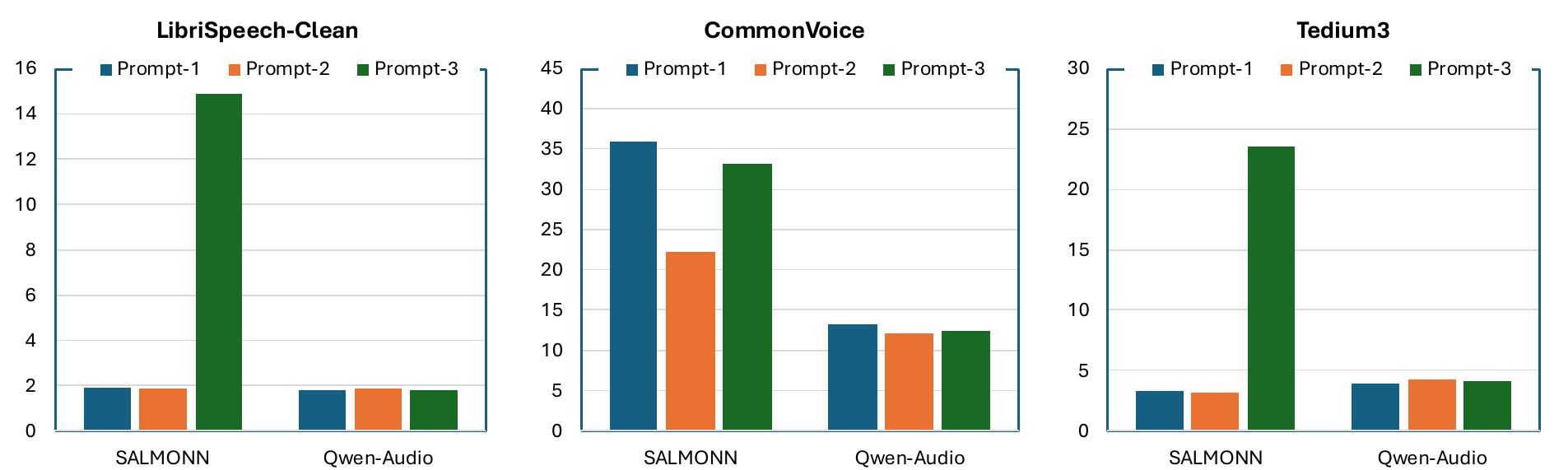}
        \caption{Sensitivity tests were conducted on three different prompts on the same datasets. Performance is measured by word error rate (WER)\textsubscript{$(\downarrow)$}. The results state that SALMONN is the least robust when faced with varying prompt templates for the same task. Prompt 1-3: \{"Turn the speech input into a text transcription", "Recognize the speech and give me the transcription", "Decode the audio and give me the written transcriptions."\}}
        \label{fig:diverse_prompt}
    \end{figure*}

        \subsubsection{Audio Scene Understanding}
    
            \textbf{Audio Question Answering (AQA)}. This task focus more on the understanding of environmental contexts. To assess the capability to follow instructions in these scenarios, we refined a test set from Clotho-AQA, retaining only the samples with high confidence levels. Simultaneously, we developed the WavCaps-QA and AudioCaps-QA datasets, each featuring over 300 diverse questions accompanied by human-verified questions and answers. Examples can be found in Table~\ref{example:aqa}.

            \noindent\textbf{Audio Captioning (AC)}. This task involves generating captions (descriptions) for an audio clip. We incorporate two popular datasets to assess this task: WavCaps and AudioCaps. The examples can be found in Table~\ref{example:ac}. \cite{banerjee2005meteor}

        \subsubsection{Voice Understanding}

            \textbf{Emotion Recognition (ER)}. Emotion is a critical paralinguistic element that can be conveyed through human speech or non-speech content~\cite{el2011survey}. It plays a vital role in making communications interpretable for listeners. Three datasets are incorporated and the examples can be found in Table~\ref{example:er}.

            \noindent\textbf{Accent Recognition (AR)}. Accent, a significant paralinguistic feature, has been overlooked in previous evaluations. We have included metadata from the VoxCeleb1 dataset and tasked the model with predicting the speaker's likely origin based on their accent. Examples can be found in Table~\ref{example:ar}.
        
            \noindent\textbf{Gender Recognition (GR)}. Gender could be recognized based on vocal characteristics. This task tests the model's sensitivity to gender-specific acoustic patterns. We have prepared two datasets and the examples can be found in Table~\ref{example:gr}.

\section{Results and Analysis}
\label{sec:results_analysis}

    \subsection{Main Results}
    
    For comparative analysis, we selected four representative AudioLLMs: SALMONN~\cite{tang2023salmonn}, Qwen-Audio~\cite{chu2023qwen}, WavLLM~\cite{hu2024wavllm} and Qwen2-Audio~\cite{chu2024qwen2}. These models are among the most capable multi-task speech-text models available, designed to manage various tasks simultaneously. Additionally, we include a cascade model named Whisper+Llama3, which processes data in a pipeline manner. First, transcriptions are extracted using the Whisper-larg-v3~\cite{huang-tsai-2023-whisper} model, and then these transcriptions along with user queries are input into the Llama-3-8B-Instuct model to generate responses. While this model cannot comprehend rich audio content, relying solely on transcriptions for context, it demonstrates strong performance in speech-intensive tasks and serves as a robust baseline.

    Table~\ref{tab:results} detailed the overall results across 26 datasets for 5 models, revealing that no single model consistently outperforms the others on all tasks. Specifically, SALMONN model is sensitive to varying instructions for ASR tasks, which indicates that it is overfitting to certain audio features and ignores proper instructions. Qwen1/2-Audio and WavLLM demonstrate robust ASR capabilities. However, all AudioLLMs struggle with long-form ASR tasks. This could be because the models are mainly fine-tuned with limited audio length which makes them hard to generalize to arbitrary context lengths. Additionally, the training and pre-training datasets for AudioLLMs are generally smaller than those used for other task-specific models like Whisper, which limits their generalizability to unnatural truncation as well.

    For speech-intensive tasks such as SQA and SI, the cascade model Whisper+Llama3 exhibits superior performance. This effectiveness stems from the Whisper model's robust recognition capabilities and Llama's strong reasoning abilities, which together efficiently handle queries primarily contained in the verbal content. In contrast, the modality fusion process in AudioLLMs may distort speech content, highlighting an area for future improvement.

    For tasks involving paralinguistic features and non-verbal sounds, AudioLLMs generally outperform cascade models, although they do not always yield satisfactory results. An exception is found in sentiment and emotion recognition tasks, where some emotions can be directly inferred from the speech semantics. Overall, cascade models struggle with understanding non-verbal content, highlighting the need for the more robust development of AudioLLMs to handle these complexities better.

    In comparing three AudioLLMs, we observed that models perform better when exposed to related tasks during training. A significant example is WavLLM, which specializes in speech content integration. It excels in SQA tasks but lacks exposure to non-spoken scenarios such as audio captioning. Consequently, it underperforms in audio scene understanding and vocal understanding tasks, demonstrating that its instruction-following capabilities are not readily generalizable without specifically aligned training samples.

    \begin{figure*}[htb]
     \centering
         \includegraphics[width=0.98\textwidth]{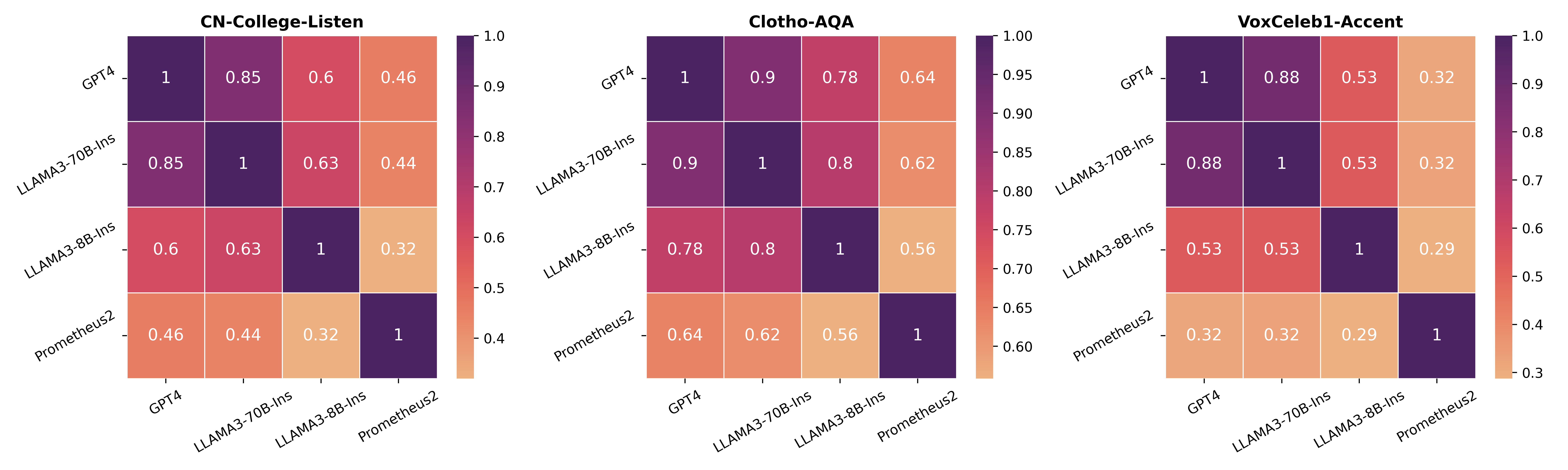}
        \caption{Correlation results for three benchmark datasets across three variations of the model-as-judge approach were analyzed using Spearman's rank correlation. The findings show that Llama-3-70B-Instruct has a very strong correlation with GPT-4, when serving as the judge.}
        \label{fig:correlation}
    \end{figure*}

    \subsection{Robustness Queries}
    
    In evaluation, we found that AudioLLMs could behave differently to different prompt templates. Even though this also applies to text-based LLMs, it is even more severe in AudioLLMs as the two modalities are not aligned in the first stage and aggressive alignment would hurt the generalizability of different prompts. \citet{hu2024wavllm} also spot such issues but only limited experimental studies are presented. Therefore, we did a more thorough experimental evaluation in this case with 2 models (SALMONN and Qwen-Audio) on 3 ASR datasets (LirbriSpeech-Clean, CommonVoice, Tedium3).

    Figure~\ref{fig:diverse_prompt} illustrates the results. We can witness that SALMONN can effectively perform ASR tasks and achieve performance comparable to the Qwen-Audio model on the LibriSpeech-Clean and Tedium3 tasks, provided the right prompts are used. However, prompt 3, "Decode the audio and give me the written transcripts," triggers the SALMONN model to conduct phoneme recognition on the LibriSpeech and Tedium3 datasets. For the CommonVoice dataset, SALMONN tends to perform speech translation on a significant number of samples, which adversely affects its performance on the Word Error Rate (WER) metric. This suggests that SALMONN may be overly tuned to speech features (tokens) and not sufficiently responsive to the prompts. In contrast, Qwen-Audio's performance remains stable across all three prompt templates. Consequently, to minimize variability and thoroughly test model robustness, we incorporate at least 20 diverse prompt templates into our evaluation framework for the tasks without diverse prompts originally.

    \subsection{Model-as-Judge Comparison}
    
    In many datasets, we use an open-ended generation style, which is considered the most natural method of interaction. However, this style introduces challenges for evaluation, as traditional metrics are not easily adaptable. To address this, we employ a model-as-judge approach to assess model responses against reference answers and questions. Yet, there are ongoing challenges in selecting appropriate models to ensure both accessibility and robustness.

    The most frequently-used judgement model is GPT-4. However, GPT-4 is closed-source, subject to ongoing updates, and often incurs costs that may not be feasible for the wider community. To explore alternatives, we examined 3 open-source models: Llama-3-8B-Instruct, Llama-3-70B-Instruct, and Prometheus-2~\cite{kim2024prometheus}. Prometheus-2 is specifically fine-tuned for response grading, while Llama-3-Instruct is a general-purpose open-source model. 
    
    To identify the most suitable open-source model for evaluation tasks, we use all samples from the following three datasets: CN-College-Listen, Clotho-AQA, and VoxCeleb1-Accent, to evaluate their replaceability of GPT-4. The three datasets cover speech understanding, audio scene understanding, and voice understanding. Initially, we get the model outputs from the SALMONN model. These model outputs, along with golden answers and questions, are then fed into the four judgment models to obtain judgment scores. We subsequently calculated the Spearman's rank correlation to compare the effectiveness of the different judgment models. The evaluation template can be found in Table~\ref{example:model_judge}. The results are detailed in Figure~\ref{fig:correlation}.

    As a result, the Llama-3-70B-Instruct model exhibits the highest correlation with GPT-4-as-a-judge (gp4-turbo-0409). The correlation scores exceeding 0.85 across all three datasets indicate a "very strong correlation" with GPT-4. We attribute this to Llama-3-70B-Instruct's robust generalizability. In contrast, Prometheus-2, though fine-tuned from the Mistral model and further adapted for scoring purposes, does not effectively compensate for the limitations of its base model. Therefore, in experiments, we will use Llama-3-70B-Instruct as the judging model (scale from 0-5 or 0-1 followed by rescaling to 100-point scale for easier interpretation), as it provides transparency and offers ease of adaptability. It is important to note that using a model-as-judge is not an ultimate solution. The issue of accurately grading free-form generations remains an unresolved challenge in NLP.

\section{Conclusion}

    In this work, we introduce the first comprehensive evaluation benchmark for AudioLLMs, named AudioBench, which includes 8 tasks and 26 datasets. Our report includes a detailed analysis of data selection, metric design, and the evaluation pipeline. We emphasize the critical factors for benchmark design and model development. We anticipate that our proposal will stimulate further progress in multimodal foundation models.

\section*{Limitations}

    First, the current AudioBench exclusively includes English datasets. However, multilingual capabilities and code-switching are crucial for comprehensive speech understanding and generation. We plan to expand the benchmark to incorporate these aspects in future iterations.

    Second, similar to text-based LLMs, evaluating free-style generation is challenging and demands robust metrics or models to serve as judges. Traditional metrics fall short in zero-shot scenarios, and assessing the correctness of output responses is complex. Thus, the development of suitable evaluation metrics, particularly for audio inputs, is crucial.

    Third, as end-to-end models, AudioLLMs typically involve large model sizes, which result in longer inference times. In this benchmark, our focus has been primarily on accuracy rather than efficiency. Moving forward, it will be important to consider inference speed and the deployment environment to comprehensively evaluate these models during the deployment process.

\section*{Acknowledgement}

    This research is supported by the National Research Foundation, Singapore, under its National Large Language Models Funding Initiative. Any opinions, findings, conclusions, or recommendations expressed in this material are those of the author(s) and do not reflect the views of the National Research Foundation, Singapore.

    We are grateful to Ziyi Xu (NUS, Singapore) and Anh Thuc Nguyen (UNC, USA) for their contributions to annotations and quality verification.

\bibliography{anthology,custom}

\appendix

\begin{table*}[thb]
\begin{adjustbox}{width=1.00\textwidth, center}
\begin{tabular}{ m{3.8cm} | m{13cm} }
\toprule

\textbf{Judgement Model} & \textbf{Template} \\  \hline 

\multirow{1}{*}{\textbf{Llama-3-70B-Instruct}} & \emph{[Reference Answer]} \\ 
\multirow{1}{*}{\&} & \emph{\{reference\}} \\
\multirow{1}{*}{\textbf{Llama-3-8B-Instruct}} &  \\
\multirow{1}{*}{\&} & \emph{[Model Answer]}  \\
\multirow{1}{*}{\textbf{GPT4}} & \emph{\{prediction\}} \\
&  \\
& \emph{[Question]}  \\
& \emph{\{question\}} \\
&  \\
& \emph{[Task]}  \\
& \emph{Rate the model's answer based on its alignment with the reference answer, focusing on accuracy and relevance to the reference provided. Please be critical on the details.}  \\
& \emph{Criteria: Assess if the model's response mirrors the reference in terms of content, accuracy, and relevance.}  \\
& \emph{Score0: The answer is completely misaligned, providing incorrect or irrelevant information compared to the reference.}  \\
& \emph{Score1: The answer shows minimal alignment, often misunderstanding or providing irrelevant details unrelated to the reference.}  \\
& \emph{Score2: The answer recognizes the topic but diverges significantly from the reference in accuracy or relevance.}  \\
& \emph{Score3: The answer aligns with the reference generally but lacks detail or precise accuracy in some aspects.}  \\
& \emph{Score4: The answer is mostly accurate and relevant, closely following the reference but could be clearer or more detailed.}  \\
& \emph{Score5: The answer is highly accurate, detailed, and matches the reference answer perfectly, capturing its essence and detail.}  \\
&  \\
& \emph{Your response should be formatted as follows:}  \\
& \emph{Explanation: (Provide a concise explanation of your rating, comparing the reference answer with the model's response. "The reference answer is [XXX], while the model's answer is [YYY]. I think ...")}  \\
& \emph{Rating: (int)"""}  \\

\hline

\textbf{Prometheus2} & \emph{"criteria": "Does the model provide accurate, relevant, and contextually appropriate responses to user inquiries?"} \\
& \emph{"score1\_description": "The model frequently fails to understand or address the core of the user's inquiries, providing inaccurate, irrelevant, or inappropriate responses."}  \\
& \emph{"score2\_description": "The model occasionally recognizes the topic of inquiry but often provides responses that are not sufficiently accurate, detailed, or contextually relevant."}  \\
& \emph{"score3\_description": "The model usually understands the question and attempts to provide a relevant answer, yet the responses may sometimes lack detail, accuracy, or context."}  \\
& \emph{"score4\_description": "The model consistently understands and appropriately addresses the questions, providing accurate and relevant responses. However, there may still be minor inaccuracies or instances where additional context could enhance clarity."}  \\
& \emph{"score5\_description": "The model excels in understanding user inquiries and consistently delivers accurate, detailed, and contextually appropriate responses that thoroughly address the user's needs."}  \\

\bottomrule

\end{tabular}
\end{adjustbox}
\caption{Template for Model-as-Judge.}
\label{example:model_judge}
\end{table*}

\section{Appendix}
\label{sec:appendix}

\subsection{Discussion on Dataset}

\begin{figure*}[t!]
	\centering
	\resizebox{1.0\textwidth}{!}{
	\begin{forest}
  for tree={
  grow=east,
  reversed=true,
  anchor=base west,
  parent anchor=east,
  child anchor=west,
  base=left,
  font=\small,
  rectangle,
  draw,
  rounded corners,align=left,
  minimum width=2.5em,
  inner xsep=4pt,
  inner ysep=1pt,
  },
  where level=1{text width=5em,fill=blue!10}{},
  where level=2{text width=5em,font=\footnotesize,fill=pink!30}{},
  where level=3{font=\footnotesize,yshift=0.26pt,fill=yellow!20}{},
  [\textbf{AudioBench} \\ Benchmark for AudioLLMs,fill=green!20
        [Automatic Speech Recognition (ASR),text width=13em
            [LibriSpeech-Clean~\cite{panayotov2015librispeech},text width=14.5em]
            [LibriSpeech-Other~\cite{panayotov2015librispeech},text width=14.5em]
            [CommonVoice~\cite{ardila-etal-2020-common},text width=13.5em]
            [PeoplesSpeech~\cite{galvez2021people},text width=13.5em]
            [Earning-21~\cite{del2021earnings},text width=13.5em]
            [Earning-22~\cite{del2022earnings},text width=13.5em]
            [GigaSpeech~\cite{chen2021gigaspeech},text width=13.5em]
            [Tedium3~\cite{rousseau2012ted},text width=13.5em]
            [Tedium3-Longform~\cite{rousseau2012ted},text width=15em]
        ]
        [Speech Question Answering (SQA),text width=12em
            [CN-College-Listen~\cite{hu2024wavllm},text width=18em]
            [SLUE-P2-SQA5~\cite{shon2022slue},text width=13.5em]
            [DREAM-TTS~\cite{sun2019dream},text width=13.5em]
            [Public-SG-SpeechQA,text width=13.5em]
        ]
        [Audio Question Answering (AQA),text width=12em          
            [Clotho-AQA~\cite{lipping2022clotho},text width=13.5em]
            [WavCaps-QA~\cite{mei2023wavcaps},text width=13.5em]
            [AudioCaps-QA~\cite{audiocaps},text width=13.5em]
        ]
        [Emotion Recognition (ER),text width=9em               
            [IEMOCAP-Emotion~\cite{busso2008iemocap},text width=13.5em]
            [MELD-Sentiment~\cite{poria-etal-2019-meld},text width=13.5em]
            [MELD-Emotion~\cite{poria-etal-2019-meld},text width=13.5em]
        ]
        [Speech Instruction (SI),text width=8em              
            [OpenHermes-Audio~\cite{OpenHermes},text width=15em]
            [ALPACA-Audio~\cite{alpaca},text width=15em]
        ]
        [Audio Captioning (AC),text width=8em              
            [WavCaps~\cite{mei2023wavcaps},text width=13.5em]
            [AudioCaps~\cite{audiocaps},text width=13.5em]
        ]
        [Accent Recognition (AR),text width=8.5em              
        [VoxCeleb-Accent~\cite{nagrani2017voxceleb},text width=13.5em]
        ]
        [Gender Recognition (GR),text width=8.5em              
            [VoxCeleb-Gender~\cite{nagrani2017voxceleb},text width=13.5em]
            [IEMOCAP-Gender~\cite{busso2008iemocap},text width=13.5em]
        ]
        [Future Tasks,text width=7em              
            [Audio Generation Tasks,text width=9em 
                [Text-to-Speech (TTS)] 
                [Voice Conversion (VC)]
                [Speech Segmentation (SS)]
            ]
            [Multilingual Understanding,text width=10em
                [Speech Translation (ST) \& others.]]
            [Multiple Audio Understanding,text width=10.5em
                [Speaker Verification (SV) \& others.]
            ]
            [etc.,text width=5em]
        ]
    ]
\end{forest}
	}
	\caption{Structure of AudioBench datasets.}
	\label{fig:appendix-main}
    \end{figure*}
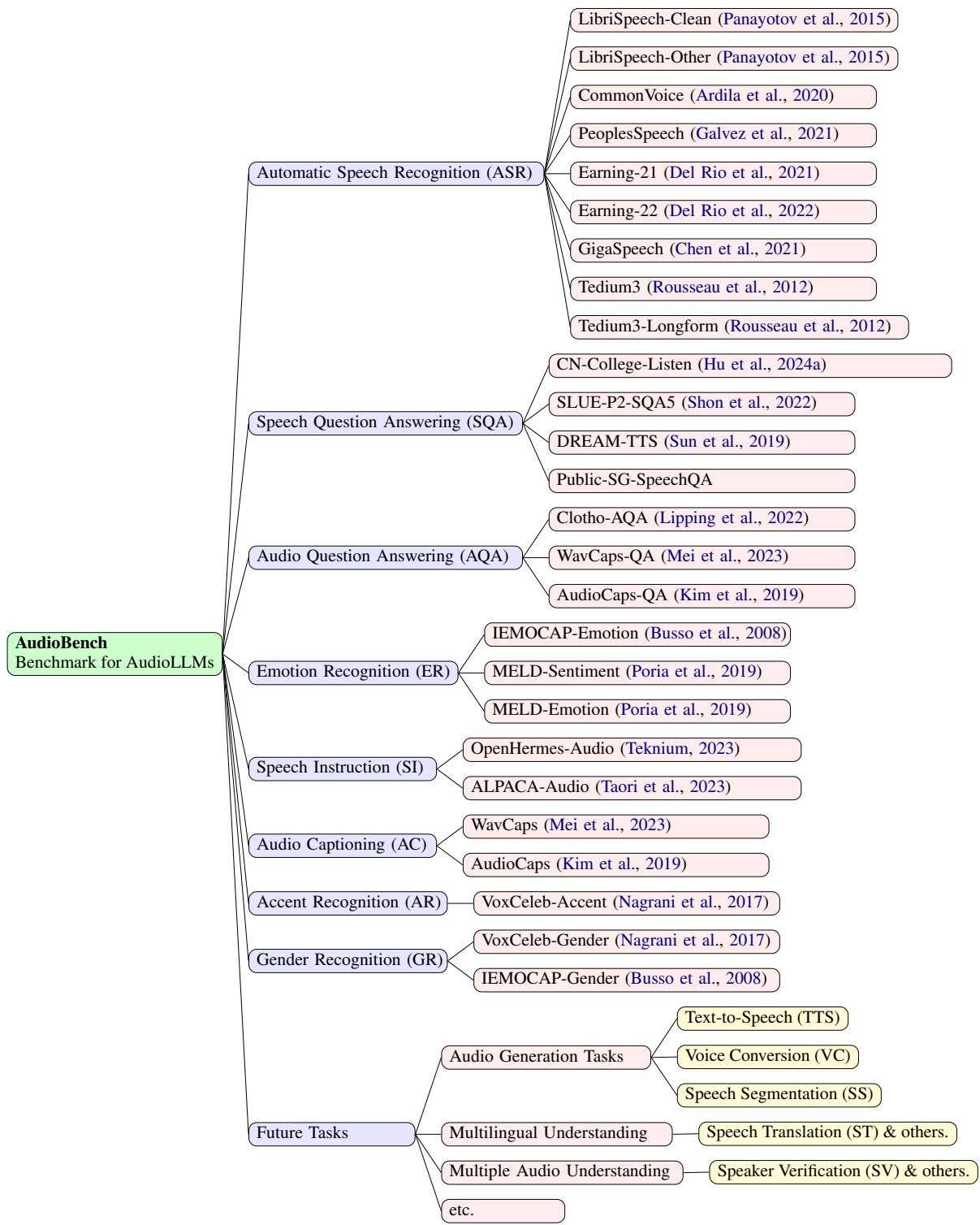

    In Figure~\ref{fig:appendix-main}, we present the structure of the tasks and datasets included in our benchmark. In this section, we aim to provide a detailed description of the newly curated datasets along with their construction processes.

    \textbf{CN-College-Listen}. This is a zero-shot testing set compiled from English comprehension questions extracted from college exam papers. Initially, we downloaded audio files and corresponding exam papers from public websites. These audio files were segmented according to the questions in the exam papers. Typically, an audio segment may relate to multiple questions depending on the sections of the exam paper. Each segmented audio is paired with its corresponding questions to formulate the test items. Originally, the questions in the exam papers are formatted as multiple-choice; however, for the application with AudioLLMs, it is more natural to frame them as open-ended questions. We retain the correct choice as the reference answer. In total, we collected 271 audio-question-answer triples and supplemented these with another 2000 questions from~\cite{hu2024wavllm} to create the final dataset. This dataset requires a precise capture of audio content, including both monologues and dialogues.

    \textbf{DREAM-TTS}. The evaluation set for spoken dialogue understanding remains limited. DREAM is a dataset featuring human-annotated questions and answers related to a dialogue. We used text-to-speech (TTS) technology to convert the original text dialogues into spoken format, adhering to gender information to closely mimic real-life scenarios. The original questions and answers were retained to create dialogue-question-answer triples.

    \textbf{Public-SG-SpeechQA}. To evaluate real speech content, we selected four videos from public speeches in Singapore. These speeches vary in length from 20 minutes to an hour, and clean transcripts are available online. We manually segmented the speeches based on topic transitions. Using the Llama-2-7B-Chat model, we generated question-and-answer pairs for each segment. These pairs underwent a subsequent round of human review to ensure they were concrete and relevant. Ultimately, we compiled 688 speech-question-answer triples, with each speech segment corresponding to multiple questions and answers.

\begin{figure*}[thb]
 \centering
     \includegraphics[width=1.00\textwidth]{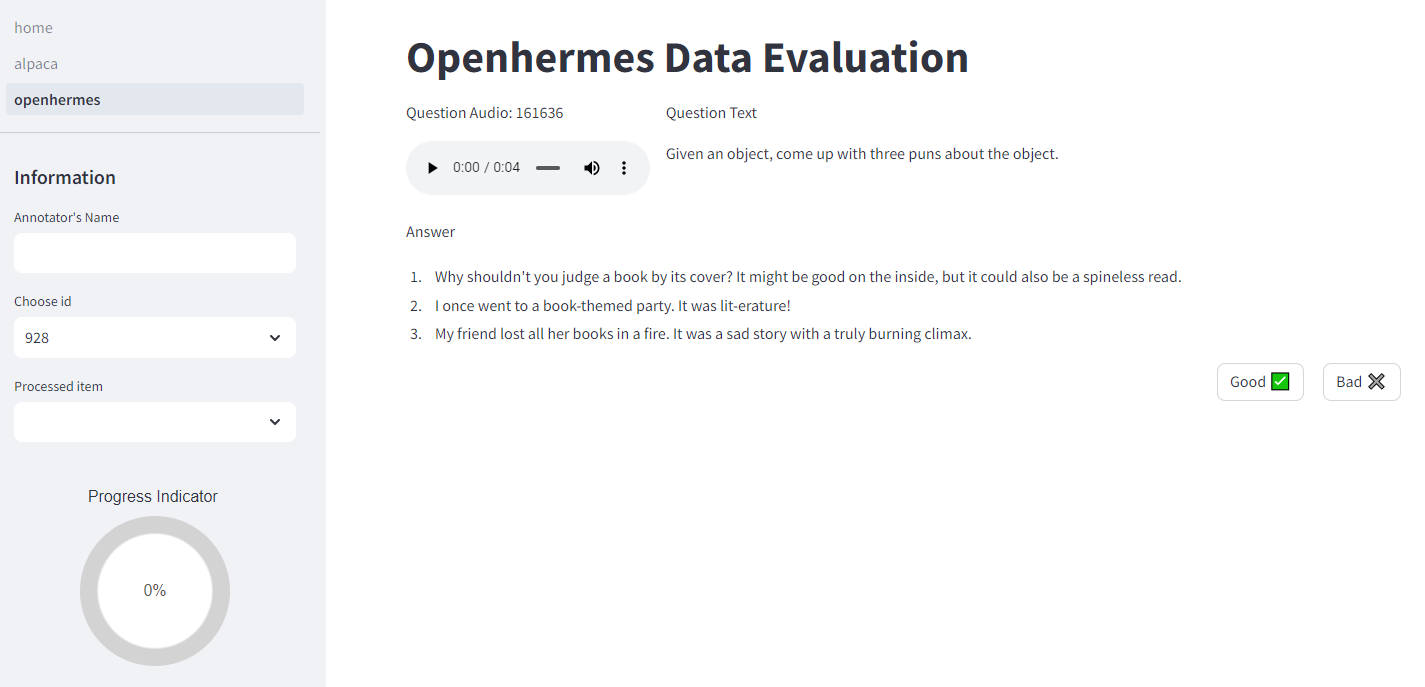}
    \caption{Our Labeling Platform for OpenHermes-Audio.}
    \label{fig:openhermes_labeling}
\end{figure*}

\begin{figure*}[thb]
 \centering
     \includegraphics[width=1.00\textwidth]{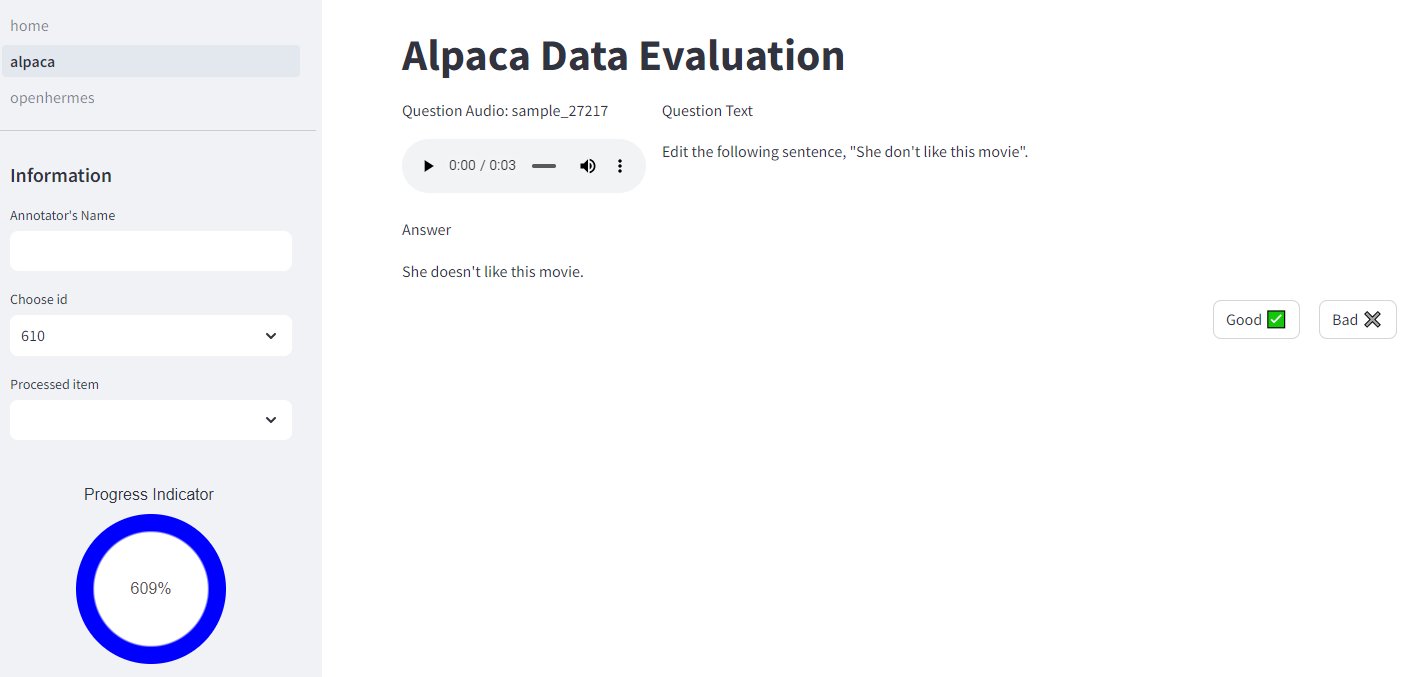}
    \caption{Our Labeling Platform for ALPACA-Audio.}
    \label{fig:alpaca_labeling}
\end{figure*}

    \textbf{OpenHermes-Audio} and \textbf{ALPACA-Audio}. Since effective speech instruction datasets are lacking, we utilized an open-source instruction tuning set designed for text-based LLMs and converted these instructions into spoken format using text-to-speech (TTS) technology. Human reviewers then verified these converted samples, filtering out approximately 90\% and retaining only 10\% for our test set. This filtering process ensures the readability, naturalness of the spoken instructions, and correctness of the reference answers. Figure~\ref{fig:openhermes_labeling} and \ref{fig:alpaca_labeling} show the annotation platform we developed using Streamlit for data annotation for these two datasets.

\begin{figure*}[thb]
 \centering
     \includegraphics[width=1.00\textwidth]{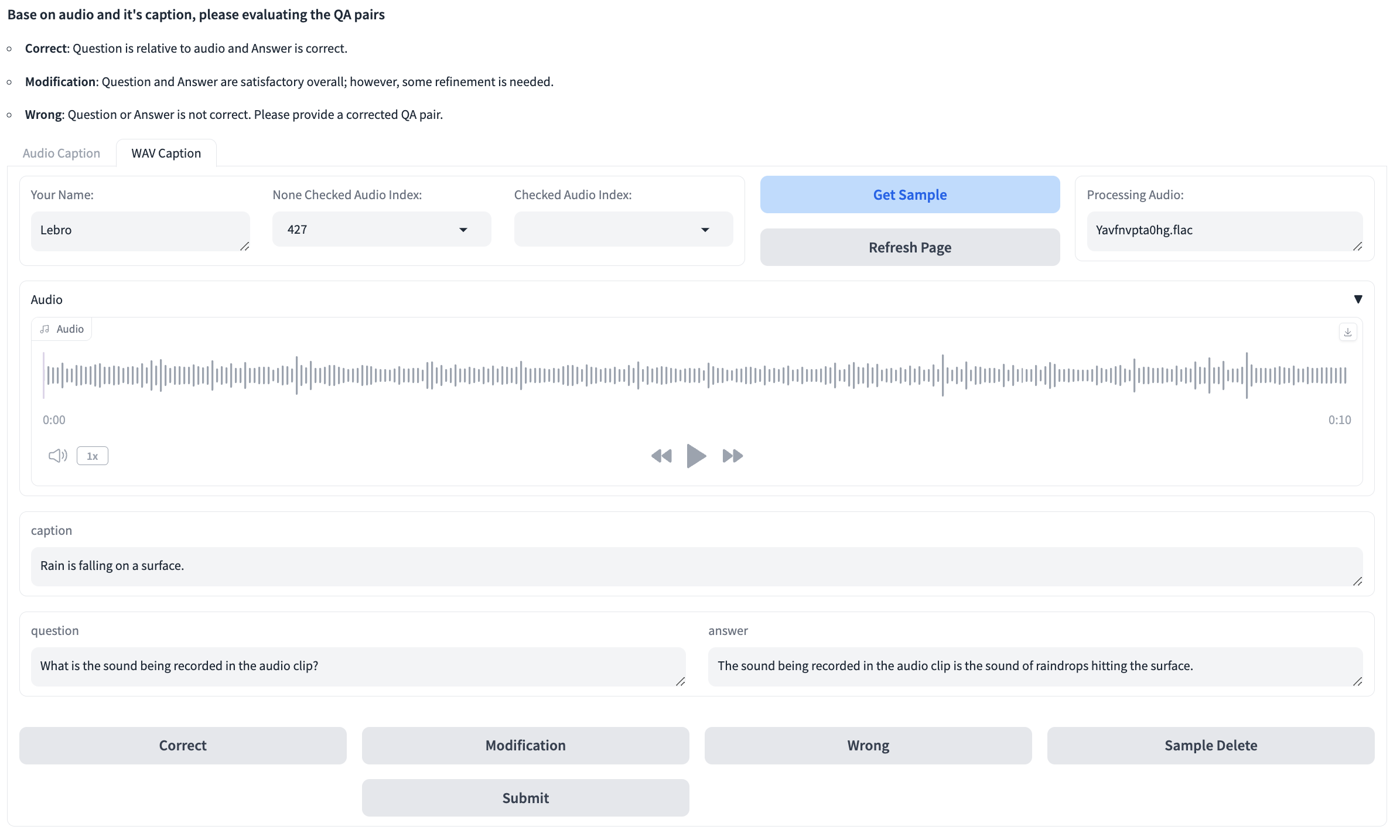}
    \caption{Our Labeling Platform for AudioCaps QA and WavCasp QA.}
    \label{fig:wav_labeling}
\end{figure*}

    \textbf{WavCaps QA} and \textbf{AudioCaps QA}. Figure~\ref{fig:wav_labeling} illustrates the data annotation platform we developed for the WavCaps and AudioCaps datasets. Initially, we used the Llama-3-8B-Instruction model to generate questions and answers from the provided captions. Each generated sample then went through another round of human annotation to refine the questions and ensure the validity of the test set. Answers were also revised as needed. Through this process, we compiled over 300 questions for each dataset to assess the models' performance effectively.

    \subsection{Model-as-Judge Template}

        Table~\ref{example:model_judge} the scoring templates used for the model-as-judge approach. The scoring range is from 0 to 5, except for Prometheus2, where the range is from 1 to 5. For some tasks that require binary judgment, the score will be either 0-incorrect or 1-correct. In practice, we rescaled all scores to a 100-point scale for easier comparison. For some tasks,

    \subsection{Comparison with AIR-Bench}

        AIR-Bench is one concurrent work with AudioBench~\cite{yang-etal-2024-air}. Even though both focus on evaluating audio-based instructions with LLMs. The selected datasets and evaluation preferences are different. First, the biggest difference comes from the coverage of datasets. In AudioBench, we proposed 6 new datasets to accommodate what is missing in the evaluation. Meantime, we have multiple ASR datasets, speech question answer datasets, and speech instruction datasets that are not covered by AIR-Bench. Second, the AudioBench evaluation toolkit takes care of prompt variants and considers the model's robustness in following instructions. Third, we present a holistic study on the choice of evaluation metrics and provide stable and affordable solutions for future benchmarking purposes. Some merits can also be seen in AIR-Bench which also includes Music datasets. Nevertheless, AudioBench servers a good foundation for evaluating AudioLLMs in three aspects: speech understanding, audio scene understanding, and voice understanding, and are expandable toolkit for more suitable scenarios.

    \subsection{Why SUPERB and Dynamic-SUPERB is not Ideal}

        For instance, a pair ($Audio$, $Query$) is presented to the model to elicit a ($Model\_Answer$), which is then compared with a reference answer for relevant tasks. This approach does not align well with previous speech benchmarks like SUPERB~\cite{yang2024large} and Dynamic-SUPERB~\cite{huang2024dynamic}. The design for SUPERB is for the evaluation of a self-supervised speech encoder where a supervised fine-tuning step is normally conducted for the evaluation. Dynamic-SUPERB can measure the instruction following tasks in a zero-shot fashion. However, it is an open collection with crowd-sourced contributions that lacks a specific focus for AudioLLMs as also discussed in \citet{yang-etal-2024-air}. Therefore, AudioBench is introduced to focus primarily on evaluating AudioLLMs, offering new datasets and evaluation pipelines to address the existing gaps.

\section{Research Outlook}
    
    \textbf{Long Audio Processing and Understanding}: The current benchmarks primarily assess the understanding capabilities of AudioLLMs using audio clips of limited duration (within minutes). However, extending the capability to process longer audio can open up broader applications such as meeting summarization and sequential event understanding. In text-based LLMs, long-sequence processing has advanced rapidly. By embedding speech content as tokens, this capability could be effectively explored.
    
    \textbf{Multi-round Query Handling}: The ability to manage multi-round queries is still limited in open-source models. Enhancing this feature would allow for more dynamic interactions where each query could involve different modalities such as images or audio, making the models more versatile in practical applications.
    
    \textbf{Multilingual Capabilities, Code-Switching, and Dialects}: Expanding the linguistic capabilities of AudioLLMs to handle multiple languages, code-switching, and various dialects is crucial. This enhancement would improve the models' applicability across diverse linguistic and cultural contexts, making them more effective for global use.
    
    \textbf{Speech Generation}: Developing more sophisticated speech generation capabilities within AudioLLMs would enable more natural and engaging human-computer interactions. This includes improving the models' ability to generate speech that not only conveys information but also mimics human-like intonations and rhythms.

\begin{table*}[tbh]
\begin{adjustbox}{width=1.00\textwidth, center}
\begin{tabular}{ m{3.1cm} | m{7cm}| m{9cm} | m{7cm}  }
\toprule

\textbf{Dataset} & \textbf{Context} & \textbf{Instruction (Example)} & \textbf{Answer} \\  \hline 

\textbf{\emph{LibriSpeech-Clean}} &
\includegraphics[width=0.025\textwidth]{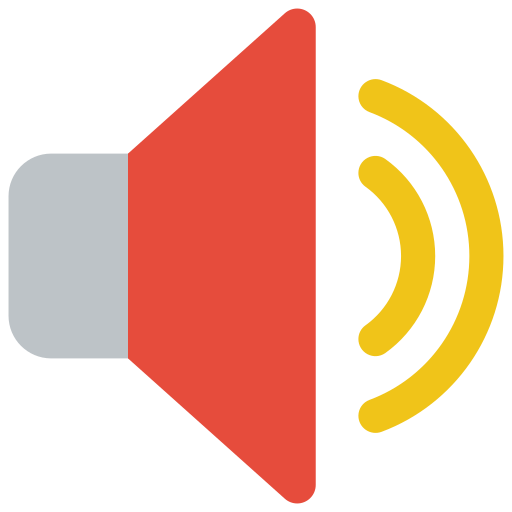}
\quad 
No, I wasn't thinking of that. &
Turn the speech input into a text transcription. &
No, I wasn't thinking of that. \\
\hline

\textbf{\emph{LibriSpeech-Other}} &
\includegraphics[width=0.025\textwidth]{audio.png}
\quad
The history of the house is plain now. &
Decode the audio and give me the written transcription. &
The history of the house is plain now. \\
\hline

\textbf{\emph{CommonVoice}} & 
\includegraphics[width=0.025\textwidth]{audio.png}
\quad
This is Jon Davis. &
Process the audio speech and provide the text output. &
This is Jon Davis. \\
\hline

\textbf{\emph{PeoplesSpeech}} & 
\includegraphics[width=0.025\textwidth]{audio.png}
\quad
that's where you have a lot of windows in ... &
Convert the audio speech into a text transcript. &
that's where you have a lot of windows in ... \\
\hline

\textbf{\emph{GigaSpeech}} & 
\includegraphics[width=0.025\textwidth]{audio.png}
\quad
I believe that fast fashion has made it possible for everyone to have access to aesthetically thoughtful clothing. &
Transform the speech into a text document. &
I believe that fast fashion has made it possible for everyone to have access to aesthetically thoughtful clothing. \\
\hline

\textbf{\emph{Earning-21}} & 
\includegraphics[width=0.025\textwidth]{audio.png}
\quad
Good morning, everyone, and welcome to the NextEra Energy Inc. and NextEra Energy Partners... &
Convert the audio speech into a text transcript. &
Good morning, everyone, and welcome to the NextEra Energy Inc. and NextEra Energy Partners...

 \\
\hline

\textbf{\emph{Earning-22}} & 
\includegraphics[width=0.025\textwidth]{audio.png}
\quad
Good day, ladies and gentlemen, and welcome to the CD Projekt Group financial results for H1 2021 conference call. Today's call... &
Transform the speech into a text document. &
Good day, ladies and gentlemen, and welcome to the CD Projekt Group financial results for H1 2021 conference call. Today's call... \\
\hline

\textbf{\emph{Tedium3}} & 
\includegraphics[width=0.025\textwidth]{audio.png}
\quad
One day, Los Angeles Times columnist Steve Lopez was walking along the streets... &
Turn the speech input into a text transcription. &
One day, Los Angeles Times columnist Steve Lopez was walking along the streets... \\
\hline

\textbf{\emph{Tedium3-Longform}} & 
\includegraphics[width=0.025\textwidth]{audio.png}
\quad
I'd like to share with you a discovery I made a few months ago while writing an article for Italian Wired. I always keep my thesaurus handy whenever... &
Process the audio speech and provide the text output. &
I'd like to share with you a discovery I made a few months ago while writing an article for Italian Wired. I always keep my thesaurus handy whenever... \\
\bottomrule

\end{tabular}
\end{adjustbox}
\caption{Examples from Automatic Speech Recognition (ASR) datasets. Capitalization normalized for display purposes.}
\label{example:asr}
\end{table*}

\begin{table*}[tbh]
\begin{adjustbox}{width=1.00\textwidth, center}
\begin{tabular}{ m{4.6cm} | m{10cm}| m{10cm} | m{4cm}  }
\toprule

\textbf{Dataset} & \textbf{Context} & \textbf{Instruction} & \textbf{Answer} \\  \hline

\textbf{\emph{CN-College-Exam-Listening}} & 
\includegraphics[width=0.025\textwidth]{audio.png}
\quad
F: Excuse me, this is the address, where do I find it? M: All right, you need a street map, here is one, and I will show you where it is. &
Question: What does the woman want to do? Choices: (A) Find a place. (B) Buy a map. (C) Get an address. &
(A) Find a place. \\
\hline

\textbf{\emph{SLUE-P2-SQA5}} & 
\includegraphics[width=0.025\textwidth]{audio.png}
\quad
1 Climate Stephens City is located in the humid subtropical climate zone ( Köppen climate classification : Cfa ) ... &
Which regions have temperate climates? &
mid-atlantic \\
\hline

\textbf{\emph{DREAM-TTS}} &
\includegraphics[width=0.025\textwidth]{audio.png}
\quad
F: The movie next Tuesday has been canceled due to a lack of interest. M: what do you mean? F: Well by last night only a few tickets been sold. &
Question: What can we conclude about the movie? Choices: (A) They want to buy the tickets for the movie. (B) The tickets for the movie were sold. (C) The movie will not be shown. &
(C) The movie will not be shown. \\
\hline

\textbf{\emph{Public-SG-SpeechQA}} & 
\includegraphics[width=0.025\textwidth]{audio.png}
\quad
Today, speaking to a roomful of economists, I am inclined to confine myself to talk about markets and price, dollars ... &
How can economics help solve complex healthcare challenges, as mentioned by the speaker? &
Economics can help solve complex healthcare ... \\
\bottomrule
\end{tabular}
\end{adjustbox}
\caption{Examples from Speech Question Answering (SQA) datasets}
\label{example:sqa}
\end{table*}

\begin{table*}[tbh]
\begin{adjustbox}{width=1.00\textwidth, center}
\begin{tabular}{ m{2.5cm} | m{6cm}| m{6cm} | m{6cm}  }
\toprule

\textbf{Dataset} & \textbf{Context} & \textbf{Instruction} & \textbf{Answer} \\  \hline 

\textbf{\emph{Clotho-AQA}} & 
\includegraphics[width=0.025\textwidth]{audio.png}
\quad
(Wave sound) &
Are there waves? &
yes \\
\hline

\textbf{\emph{WavCaps-QA}} & 
\includegraphics[width=0.025\textwidth]{audio.png}
\quad
(Electronic Music playing) &
What type of sound is being played? &
The sound being played is music. \\
\hline

\textbf{\emph{AudioCaps-QA}} & 
\includegraphics[width=0.025\textwidth]{audio.png}
\quad
(Mechanical vibration sound) &
What type of object or equipment is likely to produce a constant rattling noise and sharp vibrations? &
A loose or worn-out bolt or screw on a machine or equipment is likely to produce a constant rattling noise and sharp vibrations. \\
\bottomrule

\end{tabular}
\end{adjustbox}
\caption{Examples from Audio Question Answering (AQA) datasets}
\label{example:aqa}
\end{table*}

\begin{table*}[tbh]
\begin{adjustbox}{width=1.00\textwidth, center}
\begin{tabular}{ m{3.4cm} | m{4cm}| m{8cm} | m{6cm}  }
\toprule

\textbf{Dataset} & \textbf{Context} & \textbf{Instruction (Sample)} & \textbf{Answer} \\  \hline 

\textbf{\emph{IEMOCAP-Emotion}} & 
\includegraphics[width=0.025\textwidth]{audio.png}
\quad
Thank you.&
Can you interpret the emotions in the speaker's speech (frustration, anger, excited, neutral, happiness, surprise, sad)? &
From the speaker's speech, it seems they are in a sad state. \\
\hline

\textbf{\emph{MELD-Sentiment}} & 
\includegraphics[width=0.025\textwidth]{audio.png}
\quad
Yeah, I'm not in that.&
What sentiment signals can you hear in the speaker's speech (neutral, positive, negative)? &
From the speaker's speech, it seems they are in a neutral sentiment state. \\
\hline

\textbf{\emph{MELD-Emotion}} & 
\includegraphics[width=0.025\textwidth]{audio.png}
\quad
Yeah, I'm not in that. &
How does the speaker's speech reflect their emotional state (neutral, joy, disgust, sadness, surprise, anger, fear)? &
Based on the speaker's speech patterns, it seems like they are feeling neutral. \\
\bottomrule

\end{tabular}
\end{adjustbox}
\caption{Examples from Emotion Recognition (ER) datasets}
\label{example:er}
\end{table*}

\begin{table*}[tbh]
\begin{adjustbox}{width=1.00\textwidth, center}
\begin{tabular}{ m{3.2cm} | m{6cm}| m{6.8cm} | m{7cm}  }
\toprule

\textbf{Dataset} & \textbf{Context} & \textbf{Instruction (Sample)} & \textbf{Answer} \\  \hline 

\textbf{\emph{OpenHermes-Audio}} & 
\includegraphics[width=0.025\textwidth]{audio.png}
\quad
Pretend to be Ghost, expressing frustration to Soap that they're no closer to finding the elusive enemy leader after weeks of searching. &
Please follow the instruction in the speech &
It feels like we've been chasing ghosts, Soap. After all these weeks, we're still no closer to finding the bastard. It's bloody infuriating. \\
\hline

\textbf{\emph{ALPACA-Audio}} & 
\includegraphics[width=0.025\textwidth]{audio.png}
\quad
identify the type of the sentence: she can play guitar. &
Please follow the instruction in the speech &
The type of sentence is a declarative sentence. \\
\bottomrule

\end{tabular}
\end{adjustbox}
\caption{Examples from Speech Instruction (SI) datasets}
\label{example:si}
\end{table*}

\begin{table*}[tbh]
\begin{adjustbox}{width=1.00\textwidth, center}
\begin{tabular}{ m{1.9cm} | m{5.8cm}| m{7.5cm} | m{4cm}  }
\toprule

\textbf{Dataset} & \textbf{Context} & \textbf{Instruction (Sample)} & \textbf{Answer} \\  \hline 

\textbf{\emph{WavCaps}} & 
\includegraphics[width=0.025\textwidth]{audio.png}
\quad
(Electronic Music playing) &
Detail the ambient sounds included in the audio. &
A sound is playing. \\
\hline

\textbf{\emph{AudioCaps}} & 
\includegraphics[width=0.025\textwidth]{audio.png}
\quad
(Mechanical vibration sound)&
Describe the environmental sounds in the audio. &
Constant rattling noise and sharp vibrations \\
\bottomrule

\end{tabular}
\end{adjustbox}
\caption{Examples from Audio Captioning (AC) datasets}
\label{example:ac}
\end{table*}

\begin{table*}[tbh]
\begin{adjustbox}{width=1.00\textwidth, center}
\begin{tabular}{ m{2.8cm} | m{7cm}| m{6.8cm} | m{4cm}  }
\toprule

\textbf{Dataset} & \textbf{Context} & \textbf{Instruction (Sample)} & \textbf{Answer} \\  \hline 

\textbf{\emph{VoxCeleb-Accent}} & 
\includegraphics[width=0.025\textwidth]{audio.png}
\quad
because I was trying to get away from her, but where did you go? what did you do? &
Based on their accent, where is the speaker most likely from? &
From the audio, I guess the speaker is from USA. \\
\bottomrule

\end{tabular}
\end{adjustbox}
\caption{Examples from Accent Recognition (AR) datasets}
\label{example:ar}
\end{table*}

\begin{table*}[tbh]
\begin{adjustbox}{width=1.00\textwidth, center}
\begin{tabular}{ m{3.1cm} | m{7cm}| m{10cm} | m{5cm}}
\toprule

\textbf{Dataset} & \textbf{Context} & \textbf{Instruction (Sample)} & \textbf{Answer} \\  \hline 

\textbf{\emph{VoxCeleb-Gender}} & 
\includegraphics[width=0.025\textwidth]{audio.png}
\quad
because I was trying to get away from her, but where did you go? what did you do? &
Can you determine the gender of the speaker from the audio (Male or Female)? &
The speaker is a female. \\
\hline

\textbf{\emph{IEMOCAP-Gender}} & 
\includegraphics[width=0.025\textwidth]{audio.png}
\quad
For God's sake, Augie.  It's- Grow up, we're not going to see the grunion.&
From the audio, can you determine the speaker's gender (Male or Female)? &
Based on the auditory cues, it sounds like the speaker is a male. \\
\bottomrule

\end{tabular}
\end{adjustbox}
\caption{Examples from Gender Recognition (GR) datasets}
\label{example:gr}
\end{table*}

\clearpage

\end{document}